\documentclass{revtex4}

\usepackage{graphicx}
\def\\def\endit{\end{itemize}}
\def\begc{\begin{center}}
\def\endc{\end{center}}
\def\begl{\begin{large}}
\def\endl{\end{large}}
\def\Begl{\begin{Large}}
\def\Endl{\end{Large}}
\def\non{\nonumber \\}
\def\begeq{\begin{equation}}
\def\endeq{\end{equation}}
\def\begeqar{\begin{eqnarray}}
\def\endeqar{\end{eqnarray}}

\def\loglike{{\cal L}}

\def\mbar{{\overline m}}

\def\eth{E_{thr}}
\begin{document}
\bibliographystyle{revtex}


\title{Statistical Errors in the Measurement of Particle Thresholds}



\author{R. N. Cahn}
\email[]{rncahn@lbl.gov}
\affiliation{Lawrence Berkeley National Laboratory\\ 
1 Cyclotron Rd., Berkeley, CA 94720
}


\date{\today}

\begin{abstract}
Simple rules of thumb are derived for the precision with which s-wave and
p-wave thresholds can be determined by a series of equally spaced cross
section measurements near threshold.  Backgrounds and beam spread are ignored.
\end{abstract}

\hfill LBNL-49070

\maketitle


An important question for a linear collider is the precision with
which particle thresholds can be measured.  The thresholds in question
may have either s-wave or p-wave behavior, that is, the cross section
may rise as $p^1$ or $p^3$.  
Measurements of the production cross section near the threshold
determine particle masses quite well, but the question is how well.
The result must depend on the number of events registered, $N$, and
the interval over which the data are taken.  The beamstrahlung will
degrade the nearly monochromatic beam, but the effective beam energy
spectrum still has a delta function at the nominal energy.  The 50\%
or so of the beam that has has its energy reduced by 5\% or so simply
becomes a source of a diffuse background.  Because the highest energy
component of the beam occurs in the delta function, the sharp
threshold in the s-wave survives.  For our purposes we ignore the 
beamstrahlung altogether. This means that our estimates are too
optimistic.
We also ignore the effect of the possible widths
of the produced particles.  Detailed threshold measurements might even measure
these widths \cite{blair}.

Here we suppose the energy is set at evenly spaced discrete values and
a fixed amount of luminosity is delivered.  Let us suppose that at the
$i$th energy value the expected number of events given the cross
section with some trial values for the parameters and the delivered
luminosity is $n^{exp}_i$.  If the measured number of events is $n^{obs}_i$,
then probability of this set of measurements with these parameters is

\begeq
P=\prod_i e^{-n^{exp}_i}{{n^{exp}_i}^{n^{obs}_i}\over n^{obs}_i!}\label{eq:one}
\endeq

 Let us suppose that we 
accumulate an integrated luminosity $L$ at each of $N+1$ energies,
$E_0+(i/N)\Delta E, \ \ i=0,1,...N $. We suppose that the threshold occurs
between $E_0$ and $E_1$.  We can take the log-likelihood function $\loglike =
\ln P +{\ constant}$
to be a function of the number of observed events, $n^{obs}_i$, and the number
of expected events, $n^{exp}_i$, at each energy:
\begeq
\loglike=\sum_{i=1}^N n^{obs}_i\ln (n^{exp}_i/n^{obs}_i) - n^{exp}_i+n^{obs}_i
\endeq
where we have added the $n^{obs}_i$ term so that $\loglike$ vanishes when $n^{obs}_i=n^{exp}_i$.  The values
of $n^{exp}_i$ depend on the parameters describing the cross section.  

Now
the expected uncertainty in a parameter is given by
\begeq
\sigma^{-2}_\alpha=-{\partial^2\loglike\over\partial\alpha^2}
=
\sum_{i=1}^N{1\over n_i^{exp}}\left({\partial n_i^{exp}\over \partial \alpha}\right)^2 
\endeq
evaluated at the true value of $\alpha$.

Here we shall be interested solely in the threshold for the
process. The cross section can be written
\begeq
\sigma=\sigma_0 ((E-2m)/\eth)^\nu)
\endeq
where $2m=\eth$ is the trial value of the  threshold.  However, in order
not to mix the unknown value of $\eth$ with the parameter $m$, we
replace $\eth$ in the denominator by $E_0$,
where $E_0$ is the lowest energy setting, the one just below
threshold.  Thus $E_0$ differs from $\eth$ typically by less than a
percent.

We write
$
2\mbar=E_0+\lambda \Delta E
$ where $\mbar$ is the true mass and where $0 <\lambda< (1/N)$.  
  By hypothesis, the first non-zero contribution is for
$i=1$. 
\begeq
\sigma_m^{-2}=
\sum_{i=1}^N {4L\sigma_0\nu^2\over E_0^2} ((E_i-2\mbar)/E_0)^{\nu-2}
  = {4L\sigma_0\nu^2\over E_0^2}\epsilon^{\nu-2}\sum_{i=1}^N ({i\over N}-\lambda)^{\nu-2}
\endeq
where $\epsilon=\Delta E/E_0$.  

\begin{figure}[b]
\includegraphics[width=2.in, angle=90]{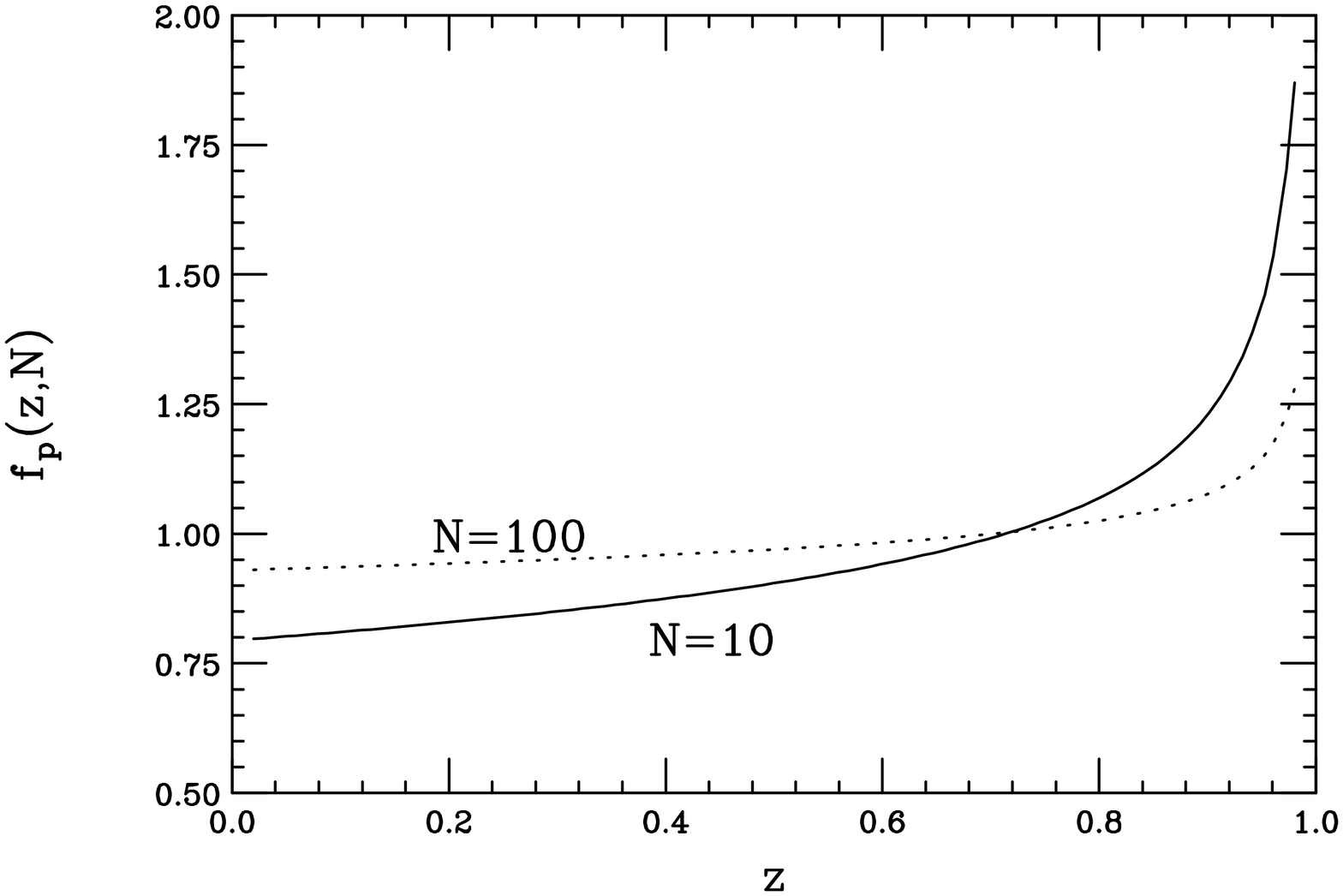}
\includegraphics[width=2.in, angle=90]{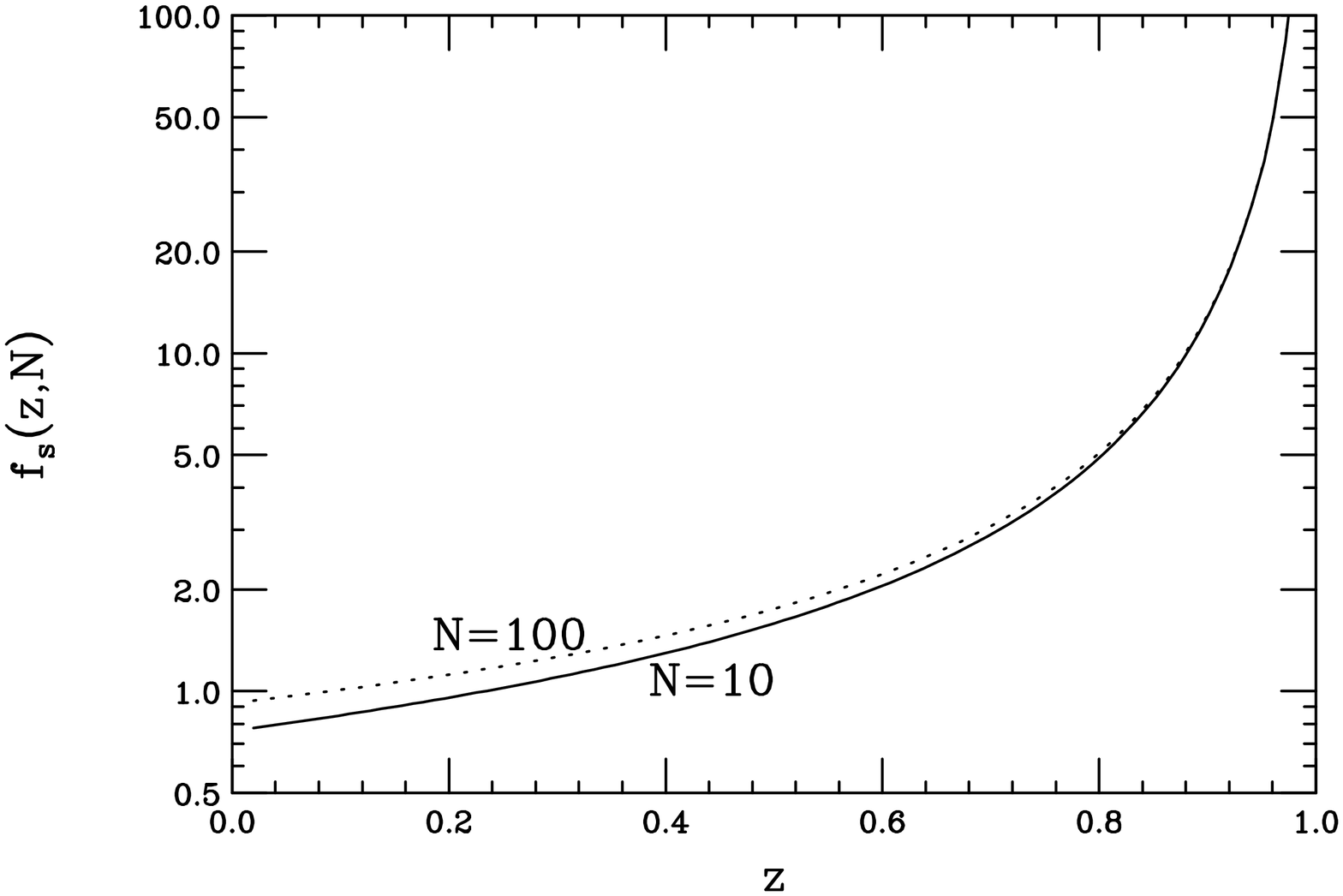}
\caption[a]{The functions $f_{p}(z,N)$ and $f_{s}(z,N)$ for two values of $N$}
\label{figone}\end{figure}

In the p-wave case $\nu=3/2$ and we have
\begeq
\sigma_m^{-2}  =  {9L\sigma_0\over E_0^2}\epsilon^{-1/2}N^{1/2}\sum_{i=1}^N ({i}-\lambda N)^{-1/2}
 =  {18L\sigma(\Delta E) N\over \Delta E^2}f_{p}(\lambda
 N,N)
\endeq
where we noted that  $\sigma_0\epsilon^{3/2}=\sigma(\Delta E)$ is the
cross section at the upper end of the $\Delta E$ interval and where
\begeq
f_{p}(z,N)={1\over 2\sqrt N}\sum_{I=1}^N (i-z)^{-1/2}
\endeq
See Fig.\ref{figone}. Thus we find
\begeq
\sigma_m= {\Delta E\over \sqrt{18 NL\sigma(\Delta E){f_{p}(\lambda
 N,N)}}}
\endeq

Martyn and Blair \cite{mandb} give an example of $e_L^-e_R^+\to
  \tilde\mu_R \tilde\mu_R$, with $\Delta E=10$ GeV,  $\sigma(\Delta
  E)=10$ fb, $N=10$, $L=10$ fb$^{-1}$.  With these values we find
\begeq
\sigma_m= {0.0745{\rm\ GeV}/\sqrt{f_{p}(\lambda
 N,N)}}
\endeq
For $\lambda=0.$ this gives 0.084 GeV, while Martyn and Blair give
0.09 GeV.
  
In the s-wave case

\begeqar
\sigma_m^{-2}  &=&  {L\sigma_0\over E_0^2}\epsilon^{-3/2}\sum_{i=1}^N ({i\over N}-\lambda)^{-3/2}\non
\endeqar
We define      
\begeq
f_{s}(z,N)={1\over \zeta(3/2)}\sum_{i=1}^N (i-z)^{-3/2}
\endeq
 where the zeta function
is defined by
$
\zeta(s)=\sum_{n=1}^\infty n^{-s}
$
so that 
$
f_{s}(0,\infty)=1
$  See Fig. \ref{figone}. We then find
\begeq
\sigma_m^{-2} 
=  {L\sigma_0 N\over \Delta
  E^2}\epsilon^{1/2}N^{1/2}{\zeta(3/2)}f_{s}(\lambda N,N)
\endeq
and
\begeqar
\sigma_m  &=& {\Delta E\over\sqrt{
    NL\sigma(\Delta E)}}N^{-1/4}[{\zeta(3/2)}f_{s}(\lambda
N,N)]^{-1/2}\non
\label{eq:s-wave}
\endeqar
where we noted that $\sigma_0\epsilon^{1/2}=\sigma(\Delta E)$ is the
cross section at the upper end of the $\Delta E$ interval. 

For the example given by Martyn and Blair \cite{mandb}, $L=10{\rm \ fb}^{-1}$,
$\sigma(\Delta E)=160$ fb, $N=10$, we find
\begeq
\sigma=0.0276[f_{s}(\lambda
N,N)]^{-1/2}
\endeq

\begin{table}\begin{center}
\begin{tabular}{|l|r|r|r|r|}\hline
Threshold (GeV)&$\lambda N $&$f_s(\lambda
N,N)$&$\sigma_m (GeV)$&$\sigma_m (GeV)$\\ 
&&&Eq.(\ref{eq:s-wave})&MC\\ 
\hline
255.0&0.0&0.77&0.030&0.030\\ \hline
255.1&0.1&0.85&0.030&0.029\\ \hline
255.5&0.5&1.59&0.022&0.021\\ \hline
255.7&0.7&2.91&0.016&0.015\\ \hline
255.9&0.9&12.8&0.007&0.007\\ \hline
\end{tabular}\label{one}\caption{Comparison of Eq.(\ref{eq:s-wave}) and toy Monte Carlo study
  with $N=10$ data points for an s-wave threshold near 255 GeV, with
  $\sigma_0=800$ fb and 10 fb$^{-1}$ at each point, corresponding to
  the study of Martyn and Blair \cite{mandb} for $e_L^-e_R^+\to
  \chi_1^-\chi_1^+$.  Martyn and Blair give an uncertainty of $0.04$
  GeV for a threshold of 255.0 GeV. As the threshold is placed closer
  and closer to the first data point, 256 GeV, the resolution steadily
  improves.  The conservative estimate is obtained by taking $\lambda N=0.$}
\end{center}\end{table}

A comparison of a toy Monte Carlo study and Eq.(\ref{eq:s-wave}) is
shown in the Table.

The functions $f_s$ and $f_p$ reflect the possible fortuitousness of
our scan.  If we get a cross section measurement just above threshold,
we get a much smaller error.  The conservative errors are obtained by
setting
$\lambda=0$ in these relations.  
We see that since the total luminosity used in the measurement is
$NL$, the p-wave measurement doesn't depend directly on the choice of
$N$, while the s-wave measurement benefits weakly from increasing $N$.

With the approximations
\begeqar
f_s(0,N)&=&1-{0.77\over \sqrt N}\non
f_p(0,N)&=&1-{0.72\over \sqrt N}
\endeqar
we have the conservative estimates
\begeqar
\sigma_p(m)&=& {\Delta E\over \sqrt{18 NL\sigma(\Delta
    E)}}}[{1+{0.36\over\sqrt N}]\non
\sigma_s(m)  &=& {\Delta E N^{-1/4}\over\sqrt{2.612
    NL\sigma(\Delta E)}}[1+{0.38\over \sqrt N}]
\endeqar

%
%

%
%


\bibliography{your bib file}

\end{document}